\documentclass{article}

\usepackage[nonatbib,final]{neurips_2020_ml4ps}

\usepackage[utf8]{inputenc} 
\usepackage[T1]{fontenc}    
\usepackage{hyperref}       
\hypersetup{
    colorlinks=true,
    linkcolor=cyan,
    filecolor=magenta,      
    urlcolor=blue,
    citecolor=blue,
}

\usepackage{url}            
\usepackage{booktabs}       
\usepackage{amsfonts}       
\usepackage{nicefrac}       
\usepackage{microtype}      
\usepackage{lipsum}
\usepackage{graphicx}
\usepackage{amsmath}
\usepackage{multirow}
\usepackage{xspace}
\usepackage{cite}
\usepackage{listings}

\usepackage{color}
\definecolor{deepblue}{rgb}{0,0,0.5}
\definecolor{deepred}{rgb}{0.6,0,0}
\definecolor{deepgreen}{rgb}{0,0.5,0}

\usepackage{array}
\newcolumntype{-}{>{\global\let\currentrowstyle\relax}}
\newcolumntype{^}{>{\currentrowstyle}}
\newcommand{\rowstyle}[1]{\gdef\currentrowstyle{#1}%
  #1\ignorespaces
}

\DeclareFixedFont{\ttb}{T1}{txtt}{bx}{n}{10} 
\DeclareFixedFont{\ttm}{T1}{txtt}{m}{n}{10}  

\newcommand\pythonstyle{\lstset{
language=Python,
basicstyle=\ttm,
otherkeywords={def},             
keywordstyle=\ttb\color{deepblue},
emph={MyClass,__init__},          
emphstyle=\ttb\color{deepred},    
stringstyle=\color{deepgreen},
frame=tb,                         
showstringspaces=false            %
}}

\lstnewenvironment{python}[1][]
{
\pythonstyle
\lstset{#1}
}
{}

\newcommand\pythoninline[1]{{\pythonstyle\lstinline!#1!}}

\renewcommand{\vec}[1]{\mathbf{#1}}

\title{Graph Generative Adversarial Networks for Sparse Data Generation in High Energy Physics}

\newcommand{\pt}{\ensuremath{p_{\mathrm{T}}}\xspace}
\newcommand{\NA}{\ensuremath{\text{---}}\xspace}

\begin{document}

\author{
   Raghav Kansal, Javier Duarte\\
  University of California San Diego \\
  La Jolla, CA 92093, USA \\
  \And
  Breno Orzari, Thiago Tomei \\
  Universidade Estadual Paulista \\
  S\~{a}o Paulo/SP - CEP 01049-010, Brazil
  \And
  Maurizio Pierini, Mary Touranakou\thanks{Also at National and Kapodistrian University of Athens, Athens, Greece.}\\
  European Organization for Nuclear Research (CERN) \\
  CH-1211 Geneva 23, Switzerland
  \And
  Jean-Roch Vlimant\\
  California Institute of Technology \\
  Pasadena, CA 91125, USA 
  \And
  Dimitrios Gunopulos\\
  National and Kapodistrian University of Athens \\
  Athens 15772, Greece
}

\maketitle

\begin{abstract}
We develop a graph generative adversarial network to generate sparse data sets like those produced at the CERN Large Hadron Collider (LHC). 
We demonstrate this approach by training on and generating sparse representations of MNIST handwritten digit images and jets of particles in proton-proton collisions like those at the LHC.
We find the model successfully generates sparse MNIST digits and particle jet data. We quantify agreement between real and generated data with a graph-based Fr\'{e}chet Inception distance, and the particle and jet feature-level 1-Wasserstein distance for the MNIST and jet datasets respectively. 
\end{abstract}


\section{Introduction}

At the CERN Large Hadron Collider (LHC), large simulated data samples are generated using Monte Carlo (MC) methods in order to translate the predictions of the standard model (SM), or beyond the SM theories, into observable detector signatures.
These samples, numbering in the billions of events, are needed in order to accurately assess the predicted yields and their associated uncertainties.
In order to achieve the highest level of accuracy possible, \textsc{GEANT4}-based simulation~\cite{Agostinelli:2002hh} is used to model the interaction of particles traversing the detector material.
However, this approach comes at a high computational cost.
At the LHC, such simulation workflows account for a large fraction of the total computing resources of the experiments, and with the planned high-luminosity upgrade, the expanded need for MC simulation may become unsustainable~\cite{Alves:2017she}. 

To accelerate simulation workflows, alternative methods based on generative deep learning models have been studied, including generative adversarial networks (GANs)~\cite{Goodfellow:2014upx,arjovsky2017wasserstein,1704.00028} and variational autoencoders (VAEs)~\cite{kingma2014auto}. 
Applications include generating particle shower patterns in calorimeters~\cite{Paganini:2017hrr,Paganini:2017dwg,Erdmann:2018jxd,Salamani:2645142,Belayneh:2019vyx,ATL-SOFT-PUB-2020-006}, particle jets~\cite{deOliveira:2017pjk,Musella:2018rdi,Carrazza:2019cnt}, event-level kinematic quantities~\cite{Otten:2019hhl,Hashemi:2019fkn,DiSipio:2019imz,Butter:2019cae}, pileup collisions~\cite{Martinez:2019jlu}, and cosmic ray showers~\cite{Erdmann:2018kuh}. 

While these studies have proven to be effective for specific high energy physics (HEP) simulation tasks, it can be both challenging and inefficient to generalize such linear and convolutional neural network architectures to a full, low-level description of collision events due to the sparsity, complexity, and irregular underlying geometry (e.g. Ref.~\cite{hgcal}) of HEP detector data. 
In this paper, we investigate a graph-based GAN to inherently account for data sparsity and any irregular geometry in the model architecture.
As noted in Ref.~\cite{GNNPP}, while graph networks have been successfully applied to classification and reconstruction tasks in HEP, they have yet to be explored for generative tasks, and this paper presents innovative work in this direction. 

As a proxy for an LHC dataset, we first consider two sparse versions of the MNIST hand-written digit dataset~\cite{MNIST}: one sparsified by hand and the other the so-called superpixels dataset~\cite{superpixels}. 
Then, we apply the same strategy to a simulated dataset of jets produced in proton-proton collisions like those occurring at the LHC~\cite{hls4mldata_30p}. 
We note that while, for convenience, we train on simulated data, for real applications this model could be trained on experimental data.


\section{Datasets}
\label{sec:data}

Our first dataset is a sparse graph representation of the MNIST dataset. 
From each image, we select the 100 highest intensity pixels as the nodes of a fully connected graph, with their feature vectors consisting of the $x$, $y$ coordinates and intensities. 
This is directly analogous to selecting the coordinates and momenta of the highest momentum particles in a jet or highest energy hits in a detector.
The second dataset, known as the MNIST superpixels dataset~\cite{superpixels}, was created by converting each MNIST image into 75 superpixels, corresponding to the nodes of a graph. 
The centers and intensities of the superpixels comprise the hidden features of the nodes.
Edge features for both datasets are chosen to be the Euclidean distance between the connected nodes. 

Finally, the third dataset~\cite{hls4ml,Coleman:2017fiq,hls4mldata_30p} consists of simulated particle jets with transverse momenta $\pt^\mathrm{jet}\approx 1$~TeV, originating from W and Z bosons, light quarks, top quarks, and gluons produced in $\sqrt{s} = 13$~TeV proton-proton collisions in an LHC-like detector.
For our application, we only consider gluon 
jets and limit the number of constituents to the 30 highest $\pt$ particles per jet (with zero-padding if there are fewer than 30).
For each particle, the following three features resulted in the best performance: the relative transverse momentum
  $\pt^{\mathrm{rel}} = \pt^\mathrm{particle}/\pt^\mathrm{jet}$ and the relative coordinates $\eta^{\mathrm{rel}}=\eta^\mathrm{particle}
  - \eta^\mathrm{jet}$ and $\phi^{\mathrm{rel}}=\phi^\mathrm{particle}
  - \phi^\mathrm{jet} \pmod{2\pi}$.
We represent each jet as a fully-connected graph with the particles as the nodes. 
A single edge feature is taken to be the $\Delta R = \sqrt{\Delta\eta^2 + \Delta\phi^2}$ between the connected particles.
For evaluation we additionally consider the jet relative mass $m^\mathrm{jet}/ \pt^\mathrm{jet}$.

\section{Graph Generator and Discriminator Architecture}
\label{sec:arch}

\begin{figure}
    \centering
    \includegraphics[width=0.8\textwidth]{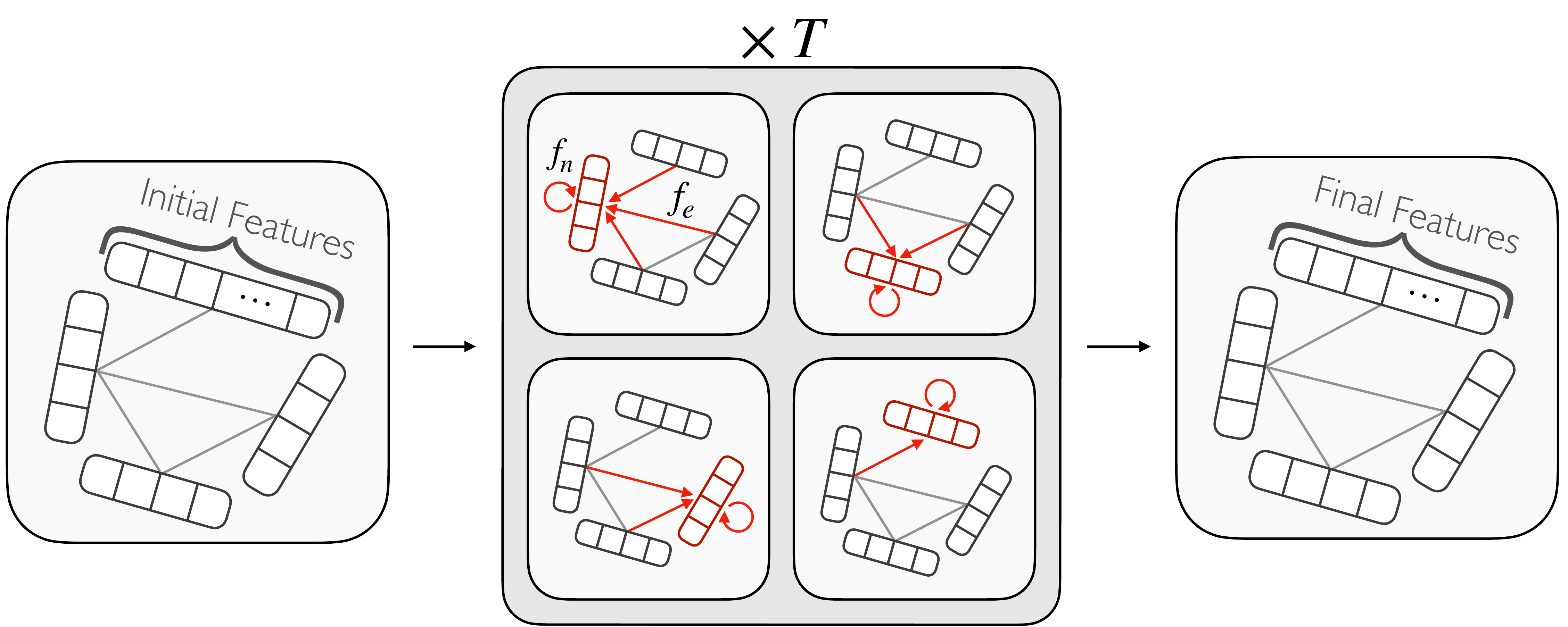}
    \caption{A message-passing neural network architecture. 
    \label{fig:mpnn}}
\end{figure}

For both the generator and discriminator we use a message-passing neural network (MPNN) architecture~\cite{MPNN}. 
For a graph $G^t = (V^t, E^t)$ after $t$ iterations of message passing ($t=0$ corresponds to the original input graph), $V^t$ a set of $N$ nodes each with its own feature vector $\vec{h}^t_v$, and $E^t$ a set of edges each with its own feature vector $\vec{e}^t_{vw}$, we define one additional iteration of message passing as
\begin{align}
        \vec{m}^{t+1}_v &= \sum_{w \in \mathcal{N}_v} f_e^{t+1}(\vec{h}^t_v, \vec{h}^t_w, \vec{e}^t_{vw}) \\
        \vec{h}^{t+1}_v &= f_n^{t+1}(\vec{h}^t_v, \vec{m}^{t+1}_v)\,,
\end{align}
where $\vec{m}^{t+1}_v$ is the aggregated message vector sent to node $v$, $\vec{h}^{t+1}_v$ is the updated hidden state of node $v$, $f_e^{t+1}$ and $f_n^{t+1}$ are arbitrary functions which are in general unique to each iteration $t$, and $\mathcal{N}_v$ is the set of nodes in the neighborhood of node $v$. 
The functions $f_e^t$ and $f_n^t$ are implemented in our case as independent multilayer perceptrons (MLPs).

The generator receives as input a graph $G^0_\mathrm{g}$ containing a set of $N$ nodes initialized with feature vectors $\vec{h}^0_v$ randomly sampled from a latent normal distribution, and then goes through $T_\mathrm{g}$ message passing iterations to output the final graph $G^{T_\mathrm{g}}_\mathrm{g}$ with new node features. 
The discriminator receives as input either a real or generated graph $G^0_\mathrm{d}$ and goes through $T_\mathrm{d}$ message passing iterations to produce a final graph $G^{T_\mathrm{d}}_\mathrm{d}$, with a single binary feature for each node classifying it as real or fake. 
This feature then is averaged over all nodes with a 50\% cutoff for the final discriminator output.

We note that a limitation of this architecture is that a particular model can only generate a fixed number of nodes and a constant graph topology. 
To overcome this we select a maximum number of nodes to produce per dataset and use zero-padding when necessary.
We leave exploring generating variable-size dynamic graph topologies to future work. 

A separate optimization is performed for every task to choose the hyperparameters $T_\mathrm{g}$, $T_\mathrm{d}$, hidden node feature size $|\vec{h}^{t}_v|$, and the number of layers and neurons in each layer of each $f_e^t$ and $f_n^t$ network.
A different model is optimized for each MNIST digit, in analogy with the HEP use case, in which different generator settings are chosen for generating different physics processes.
A variety of architectures experimented with, out of which an MPNN for both the generator and discriminator was most successful, are discussed in Appendix~\ref{app:experiments}.

\section{Training}
\label{sec:training}

We use the least squares loss function~\cite{LSGAN} and the RMSProp optimizer with a learning rate of $10^{-5}$ for the discriminator and $3\times 10^{-5}$ for the generator~\cite{TTUR}, except for the superpixel digits `2', `4', and `9' where a learning rate of $10^{-5}$ for both the generator and discriminator had better performance. 
We use LeakyReLU activations (with negative slope coefficient 0.2) for all intermediate layers, and tanh and sigmoid activations for the final outputs of the generator and discriminator respectively. 
We attempted discriminator regularization to alleviate mode collapse via dropout~\cite{dropout}, batch normalization~\cite{batchnorm}, a gradient penalty~\cite{wgangp}, spectral normalization~\cite{spectralnorm}, adaptive competitive gradient descent~\cite{acgd} and data augmentation of real and generated graphs before the discriminator~\cite{karras_2020, tran_2020, zhao_2020}.
Apart from dropout (with fraction $0.5$), none of these demonstrated significant improvement with respect to mode dropping or graph quality.

\subsection{Evaluation}
\label{sec:eval}

For model evaluation and optimization, as well as a quantitative benchmark on these datasets for comparison, we propose a graph-based Fr\'{e}chet Inception distance (FID)~\cite{TTUR} inspired metric for the MNIST datasets, and the 1-Wasserstein distance ($W_1$) for the jets dataset as in Ref.~\cite{deOliveira:2017pjk,Lu:2020npg}.
The two metrics differ for reasons explained below.


Traditionally, the FID metric is used on image datasets, using the pre-trained Inception-v3 image classifier network. 
It compares the statistics of the outputs of a single layer of this network between generated and real samples, and has been shown to be a consistent measure of similarity between generated and real samples in terms of both quality and diversity. 
To adapt FID to graph datasets, we use the MoNet model of Ref.~\cite{superpixels} as the pre-trained classifier,
which can be found at Ref.~\cite{graph_gan_repo},
and calculate what we call the graph Fr\'{e}chet distance (GFD):
\begin{equation}
    \mathrm{GFD} = ||\vec{\mu}_\mathrm{r} - \vec{\mu}_\mathrm{g}||^2 + \mathrm{Tr}(\vec{\Sigma}_\mathrm{r} + \vec{\Sigma}_\mathrm{g} - 2(\vec{\Sigma}_\mathrm{r}\vec{\Sigma}_\mathrm{g})^{1/2})\,,
\end{equation} 
where $\vec{\mu}_\mathrm{r}$ ($\vec{\mu}_\mathrm{g}$) is the vector of means of the activation function outputs of the first fully connected layer in the pre-trained MoNet model for real (generated) images and $\vec{\Sigma}_\mathrm{r}$ ($\vec{\Sigma}_\mathrm{g}$) is the corresponding covariance matrix.


To evaluate the performance on the jet dataset, we calculate directly $W_1$ between the distributions of the three particle-level features and the jet $m/\pt$ in the real and generated samples. 
Unlike for MNIST, these quantities correspond to meaningful physical observables hence measuring $W_1$ between their distributions is a more desirable metric than the GFD.
We use bootstrapping to calculate a baseline $W_1$ between samples within the real dataset alone, taking $P$ pairs of random sets of $N$ jets and calculating $W_1$ between the distributions of each pair. 
For three combinations of $P$ and $N$, the means and standard deviations are shown in Table~\ref{tab:1Wmeans}.
The $W_1$ values between the real and generated distributions are similarly calculated by generating $P$ sets of $N$ jets each and comparing them to $P$ random sets of $N$ real jets.

\section{Results}
\label{sec:results}

\begin{figure}[h]
    \centering
    \includegraphics[width=\textwidth]{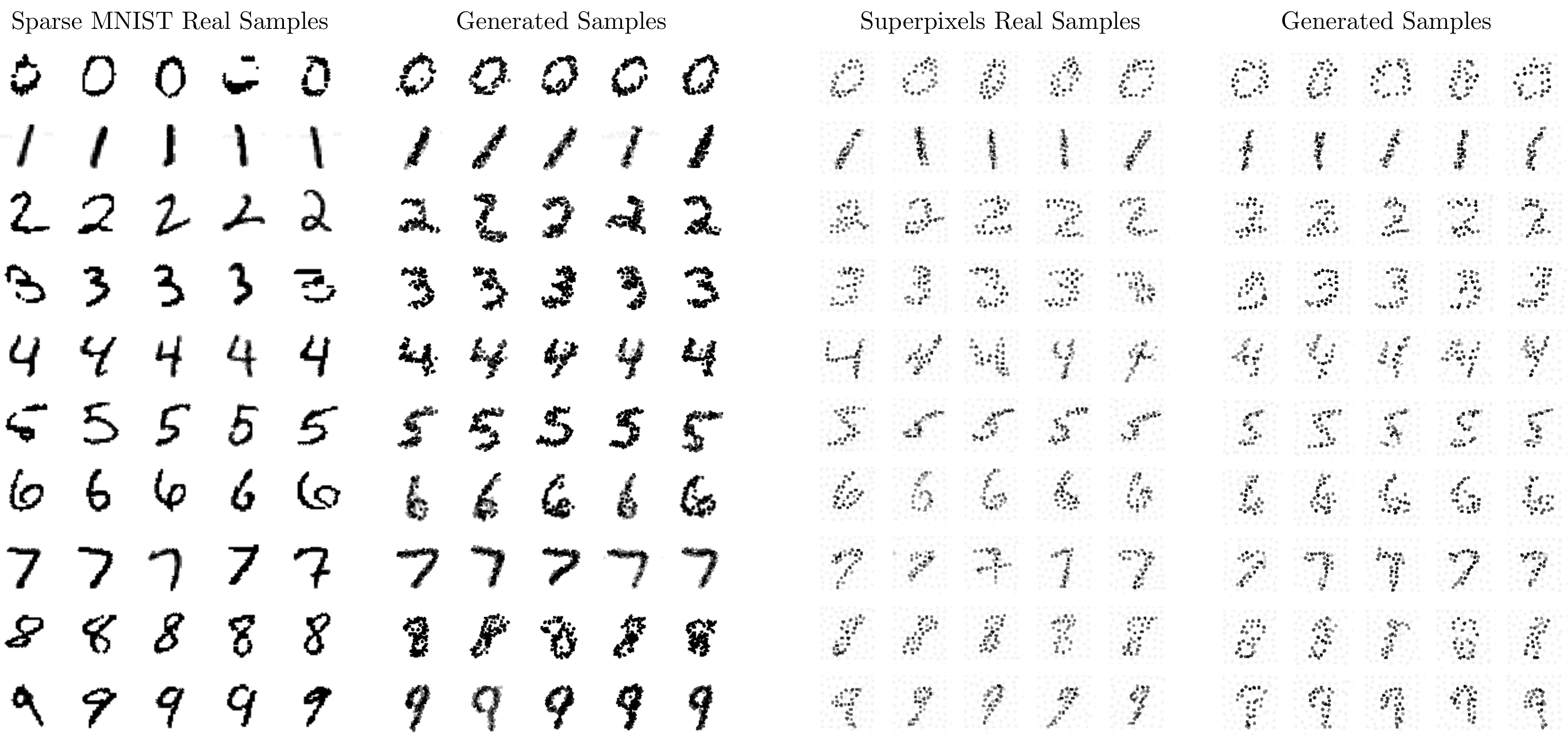}
    \caption{Samples from our sparse MNIST dataset (far left) compared to  samples from our graph GAN (center left). 
    Samples from the MNIST superpixels dataset (center right) compared to samples from our graph GAN (far right).\label{fig:real_gen}}
\end{figure}

For the MNIST-derived datasets, we optimized the hyperparameters of our model using our GFD metric. 
A sample of hyperparameter settings we tested with their corresponding GFD scores for all 10 digits can be seen in Appendix~\ref{app:hyperop}.
Based on this optimization, the final hyperparameters chosen for the three datasets as listed in Table~\ref{tab:all_hp}.
Fig.~\ref{fig:real_gen} (left) shows a comparison between real and generated digits for the sparse MNIST dataset.
The generator is able to reproduce all 10 digits successfully with high accuracy and little evidence of mode dropping.
Similarly, Fig.~\ref{fig:real_gen} (right) compares real and generated digits for the MNIST superpixel dataset.
Again, we can see that the model successfully reproduces the real samples, though there is some evidence of mode dropping, particularly with the more complex digits and rarer modes.
We leave exploring this issue further to future work.
The average of our best GFD scores across all 10 digits is 0.52 and 0.30 for the Sparse MNIST and superpixels dataset respectively.

\begin{table}[h]
\centering
\caption{Optimized hyperparameters for each dataset.
\label{tab:all_hp}}
\resizebox{\textwidth}{!}{
\begin{tabular}{c|c||c|c|c|c|c|c|c|c|c|c|c}
    \multirow{2}{*}{Dataset} & \multirow{2}{*}{Digits} & \multirow{2}{*}{$T_\mathrm{g}$} & \multirow{2}{*}{$T_\mathrm{d}$} & \multicolumn{4}{c|}{$f_e$ (Neurons per layer)} & \multicolumn{4}{c|}{$f_n$ (Neurons per layer)} & \multirow{2}{*}{$|\vec{h}^{t}_v|$}\\ \cline{5-12}
    & & & & In & 1 & 2 & Out & In & 1 & 2 & Out & \\ \hline\hline
    \multirow{2}{*}{Sparse MNIST} & 2 3 4 5 7 & 1 & 1 & 65 & 64 & \NA & 128 & 160 & 256 & 256 & 32 & 32\\
    & 0 1 6 8 9 & 1 & 1 & 65 & 96 & 160 & 192 & 224 & 256 & 256 & 32 & 32\\
    Superpixels & All & 2 & 2 & 65 & 64 & \NA & 128 & 160 & 256 & 256 & 32 & 32\\
    Jet & \NA & 2 & 2 & 65 & 96 & 160 & 192 & 224 & 256 & 256 & 32 & 32
\end{tabular}}
\end{table}

\begin{figure}[t]
    \centering
    \includegraphics[width=\textwidth]{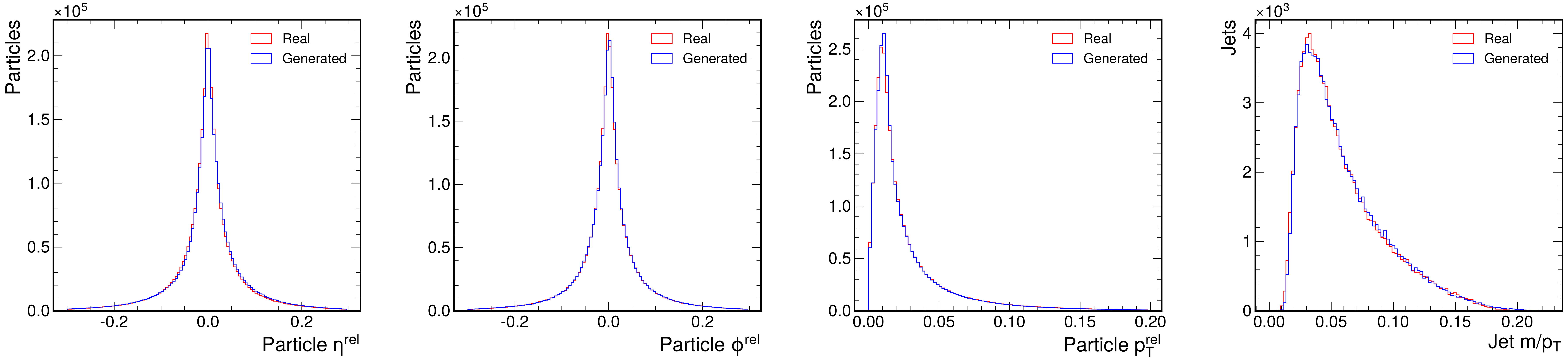}
    \caption{Distributions of the particle  $\eta^\mathrm{rel}$, $\phi^\mathrm{rel}$, and $\pt^\mathrm{rel}$, and jet $m/\pt $ for 100,000 real and generated 
    jets. 
    \label{fig:jets}}
\end{figure}

\begin{table}[htpb]
\centering
\caption{Mean $W_1$ values and standard deviations between particle-level distributions of $\eta^\mathrm{rel}$, $\phi^\mathrm{rel}$, and $\pt^\mathrm{rel}$, and the jet-level distribution of $m/\pt$ derived from comparing randomly selected sets of $N$ real jets and from comparing sets of random $N$ real and $N$ generated jets. 
This comparison is repeated $P$ times to derive a mean and standard deviation.\label{tab:1Wmeans}}
\resizebox{\textwidth}{!}{
\begin{tabular}{r|r||c|c|c|c||c|c|c|c}
    \multirow{3}{*}{$N$} & \multirow{3}{*}{$P$} & \multicolumn{8}{c}{$W_1$ mean $\pm$ standard deviation $(\times 10^{-3})$}\\
    & & \multicolumn{4}{c||}{Pairs of real distributions} & \multicolumn{4}{c}{Real and generated distributions} \\\hline
    & & $\eta^\mathrm{rel}$ & $\phi^\mathrm{rel}$ & $\pt^\mathrm{rel}$ & Jet $m/\pt$ & $\eta^\mathrm{rel}$ & $\phi^\mathrm{rel}$ & $\pt^\mathrm{rel}$ & Jet $m/\pt$ \\ \hline
    100 & 1,000 & $6 \pm 2$ & $6 \pm 2$ & $1.4 \pm 0.5$ & $6 \pm 2$ &
    $5 \pm 2$ & $11 \pm 4$ & $2 \pm 1$ & $6 \pm 2$ \\
    1,000 & 100 & $1.8 \pm 0.1$ & $1.7 \pm 0.1$ & $0.47 \pm 0.02$ & $1.9 \pm 0.7$ & 
    $2.4 \pm 0.6$ & $3 \pm 1$ & $0.7 \pm 0.2$ & $2.2 \pm 0.9$ \\
    10,000 & 10 & $0.5 \pm 0.1$ & $0.5 \pm 0.1$ & $0.11 \pm 0.02$ & $0.5 \pm 0.1$ &
    $2.2 \pm 0.2$ & $1.3 \pm 0.4$ & $0.51 \pm 0.06$ & $1.1 \pm 0.5$ \\
\end{tabular}}
\end{table}

Our results on the jet dataset using our message-passing architecture show excellent agreement, both qualitatively and quantitatively using $W_1$. 
Example generated and real distributions of particle $\eta^\mathrm{rel}$, $\phi^\mathrm{rel}$, and $\pt^\mathrm{rel}$ and jet $m/\pt$ are shown in Fig.~\ref{fig:jets} for 100,000 jets.
The mean $W_1$ values between real and generated jet distributions are presented in Table~\ref{tab:1Wmeans}.
For samples of 100 jets, the mean $W_1$ values (between real and generated jet samples) agree with the expected ones (between real jet samples) within one standard deviation, but this is not the case when the jet sample size is increased to 1,000 or 10,000.
Thus, while the generator has sufficient fidelity for smaller sample sizes, there is room for improvement for larger ones. 
We also note that there is little evidence of mode collapse with this dataset because we can see that the entire distributions, including rarer data samples in the tails, are reproduced with high accuracy. 

\section{Summary}
\label{sec:summary}

We have presented a novel architecture for generating graphs using a generative adversarial network based on a message-passing neural network, which we successfully apply to two MNIST-derived graph datasets as well as an LHC jet dataset. 
This architecture works efficiently with sparse data and inherently adapts to any underlying geometry.
We find the model generates realistic MNIST graph data albeit with some evidence of mode dropping, which we quantify with our graph Fr\'{e}chet distance.
For the jet dataset, we measure the quality of the generator using a metric based on the 1-Wasserstein distance and find high accuracy for smaller sample sizes.
The application of our model to a high energy physics dataset demonstrates its flexibility, and indicates this approach may be readily used for fast simulation of a variety of scientific datasets, including sensor-level data in high granularity calorimeters. 
%

\section*{Broader Impacts}
Physics experiments needing to generate large simulated datasets may benefit from this work. 
If this type of algorithm is used by experiments to produce such datasets, to produce datasets, it may reduce the computational cost of running the experiment.
At the same time, if the fidelity of the algorithm is not as high as desired, it may result in suboptimal or inaccurate scientific results.
Other beneficiaries of this work may include any group with a need to generate graph-based datasets following some realistic patterns.

\begin{ack}
This work was supported by the European Research Council (ERC) under the European Union's Horizon 2020 research and innovation program (Grant Agreement No. 772369).
 R.~K. was partially supported by an IRIS-HEP fellowship through the U.S. National Science Foundation (NSF) under Cooperative Agreement OAC-1836650.
J.~D. is supported by the U.S. Department of Energy (DOE), Office of Science, Office of High Energy Physics Early Career Research program under Award No. DE-SC0021187.
B. O. was partially supported by grants \#2018/01398-1 and \#2019/16401-0, São Paulo Research Foundation (FAPESP).
J-R.~V. is partially supported by the European Research Council (ERC) under the European Union's Horizon 2020 research and innovation program (Grant Agreement No. 772369) and by the U.S. DOE, Office of Science, Office of High Energy Physics under Award No. DE-SC0011925, DE-SC0019227, and DE-AC02-07CH11359.
This work was performed using the Pacific Research Platform Nautilus HyperCluster supported by NSF awards CNS-1730158, ACI-1540112, ACI-1541349, OAC-1826967, the University of California Office of the President, and the University of California San Diego's California Institute for Telecommunications and Information Technology/Qualcomm Institute. 
Thanks to CENIC for the 100~Gpbs networks.
Also, thanks to Kwonjoon Lee for valuable discussions about GAN training.
\end{ack}

\appendix

%

\section*{Appendices}

\section{Architecture Experiments}
\label{app:experiments}
We experimented with multiple GAN architectures for producing graphs. 
This included standard MLP and CNN generators and discriminators, which were predictably unsuccessful due to the architectures not being permutation invariant. 
However, despite this limitation, a CNN classifier achieved $>90$\% accuracy on our sparse MNIST dataset. 

\begin{figure}[htpb]
    \centering
    \includegraphics[width=\textwidth]{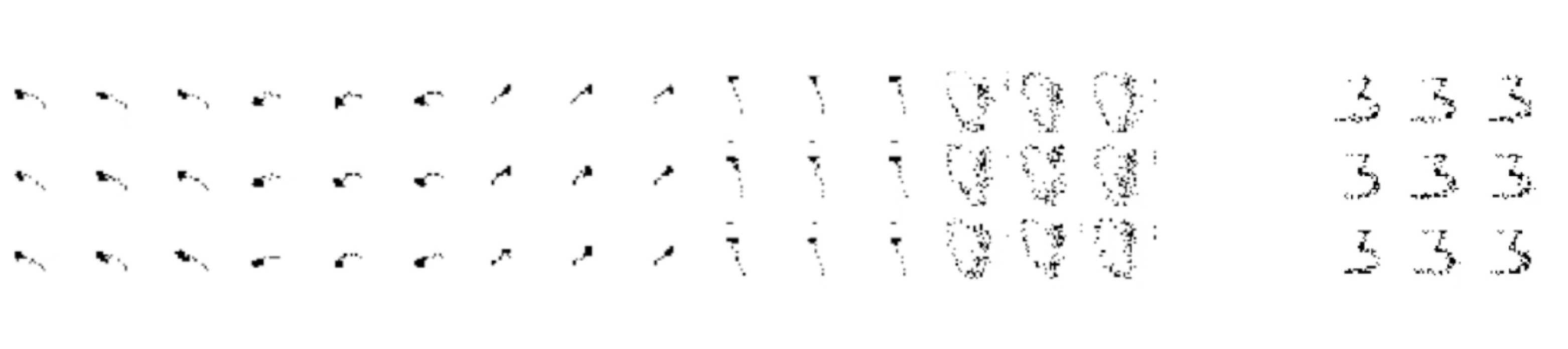}
    \caption{Samples from an RNN generator with a CNN discriminator (left), which exhibit some structure, but no coherent digits. 
    Samples from an MPNN generator with a CNN discriminator (right), which shows a successful output but displays mode collapse. \label{fig:experiments}}
\end{figure}

A better architecture we attempted was using a gated recurrent unit (GRU) based recurrent neural network (RNN) as our generator together with a CNN discriminator because of its success as a classifier. 
The RNN received as input a random sample from our latent space and iteratively output each nodes' features in sequence. 
While there was evidence of this model learning some graph structure in Fig.~\ref{fig:experiments} (left), it was not able to reproduce digits. 
Nonetheless, this model has the desirable ability to produce graphs with arbitrary numbers of nodes and future research could explore this further. 

The MPNN generator was a clear improvement and was able to successfully reproduce graphs from our dataset. 
We tested a CNN discriminator initially, and the GAN produced high-quality outputs, as seen in Fig.~\ref{fig:experiments} (right). However, training was difficult and inconsistent, and there was clear evidence of mode collapse. 
Replacing the CNN with an MPNN improved both these aspects.

With the MPNN, we experimented with an additional step in the discriminator after the message-passing iterations to produce the final classification output
\begin{equation}
    f_{nd}\left(\sum_{v \in V^{T_\mathrm{d}}_\mathrm{d}} \vec{h}^{T_\mathrm{d}}_v\right)\,,
\end{equation}
where $f_{nd}$ is implemented as an MLP. 
This allows the discriminator to take a holistic look at the graph instead of classifying on a per node basis. 
However, empirically we found that this addition marginally decreased performance so was not used in the final architecture.

For edge features, we tested the absolute Euclidean distance and the vector displacement in Cartesian and polar coordinates between node positions, as well as the difference in node intensities. 
The Euclidean distance alone was the most effective. 
We also investigated fully connected graphs versus edge connections only within a local neighborhood as in Ref.~\cite{superpixels} and the performance was comparable.

\section{Hyperparameter Optimization}
\label{app:hyperop}

A characteristic sample of hyperparameter combinations we tested for the sparse MNIST dataset digit `3' can be seen in Table~\ref{tab:sm_hp_opt}. 
A similar sample for the superpixels MNIST dataset digit `3' can be seen in Table~\ref{tab:sp_hp_opt}. 
Based on this hyperparameter optimization, the final settings for the MNIST datasets were chosen as shown in Table~\ref{tab:all_hp}.
For the jet dataset, we chose one of the hyperparameter settings optimized for the MNIST datasets and found it to be effective.

\begin{table}[htpb]
\centering
\caption{Sample of hyperparameter combinations for the sparse MNIST dataset with their respective GFD scores for digit `3.' 
The selected combination is in bold.\label{tab:sm_hp_opt}}
\begin{tabular}{-c|^c|^c|^c|^c|^c|^c|^c|^c|^c|^c|^c||^c}
    \multirow{2}{*}{$T_\mathrm{g}$} & \multirow{2}{*}{$T_\mathrm{d}$} & \multicolumn{4}{c|}{$f_e$ (Neurons per layer)} & \multicolumn{5}{c|}{$f_n$ (Neurons per layer)} &  \multirow{2}{*}{$|\vec{h}^{t}_v|$} & \multirow{2}{*}{GFD}\\ \cline{3-11}
    & & In & 1 & 2 & Out & In & 1 & 2 & 3 & Out & &\\ \hline\hline
    \rowstyle{\bfseries} 
    1 & 1 & 65 & 64 & \NA & 128 & 160 & 256 & 256 & \NA & 32 & 32 & 0.42\\
    1 & 1 & 65 & 64 & \NA & 128 & 160 & 256 & 256 & 256 & 32 & 32 & 0.50\\
    1 & 1 & 65 & 92 & 160 & 192 & 224 & 256 & 256 & 256 & 32 & 32 & 0.92\\
    2 & 1 & 65 & 64 & \NA & 128 & 160 & 256 & 256 & \NA & 32 & 32 & 0.80\\
    1 & 2 & 65 & 64 & \NA & 128 & 160 & 256 & 256 & \NA & 32 & 32 & 1.48\\
\end{tabular}
\end{table}


\begin{table}[htpb]
\centering
\caption{Sample of hyperparameter combinations for the MNIST superpixels dataset with their respective GFD scores for digit '3'.
\label{tab:sp_hp_opt}}
\begin{tabular}{-c|^c|^c|^c|^c|^c|^c|^c|^c|^c|^c|^c||^c}
    \multirow{2}{*}{$T_\mathrm{g}$} & \multirow{2}{*}{$T_\mathrm{d}$} & \multicolumn{4}{c|}{$f_e$ (Neurons per layer)} & \multicolumn{5}{c|}{$f_n$ (Neurons per layer)} &  \multirow{2}{*}{$|\vec{h}^{t}_v|$} & \multirow{2}{*}{GFD}\\ \cline{3-11}
    & & In & 1 & 2 & Out & In & 1 & 2 & 3 & Out & &\\ \hline\hline
    1 & 1 & 65 & 64 & \NA & 128 & 160 & 256 & 256 & \NA & 32 & 32 & 2.48\\
    1 & 2 & 65 & 64 & \NA & 128 & 160 & 256 & 256 & \NA & 32 & 32 & 1.93\\
    2 & 1 & 65 & 64 & \NA & 128 & 160 & 256 & 256 & \NA & 32 & 32 & 1.18\\
    \rowstyle{\bfseries} 
    2 & 2 & 65 & 64 & \NA & 128 & 160 & 256 & 256 & \NA & 32 & 32 & 0.22\\
    2 & 2 & 65 & 64 & \NA & 128 & 160 & 256 & \NA & \NA & 32 & 32 & 2.37\\
    2 & 2 & 65 & 64 & \NA & 128 & 160 & 256 & 256 & 256 & 32 & 32 & 0.44\\
    2 & 2 & 65 & 92 & 160 & 192 & 160 & 256 & 256 & \NA & 32 & 32 & 0.31\\
    2 & 2 & 17 & 64 & \NA & 128 & 136 & 256 & 256 & \NA & 8 & 8 & 1.08\\
    2 & 2 & 257 & 92 & 160 & 192 & 320 & 256 & 256 & \NA & 128 & 128 & 0.38\\
    3 & 3 & 65 & 92 & 160 & 192 & 160 & 256 & 256 & \NA & 32 & 32 & 0.62\\
\end{tabular}
\end{table}



\clearpage
\bibliographystyle{lucas_unsrt}
\bibliography{bibliography}

\end{document}